\pdfoutput=1 

\documentclass[a4paper, conference]{IEEEtran}

\usepackage[numbers]{natbib}
\usepackage[shortcuts,acronym]{glossaries}
\makenoidxglossaries
\newacronym{dpd}{DPD}{Double Pulsed Direct Method}
\newacronym{rssi}{RSSI}{Received Signal Strength Indication}
\newacronym{tof}{ToF}{Time of Flight}
\newacronym{ftm}{FTM}{Fine Time Measurement}
\newacronym{toa}{ToA}{Time of Arrival}
\newacronym{tdoa}{TDoA}{Time Difference of Arrival}
\newacronym{twr}{TWR}{Two Way Ranging}
\newacronym{sds-twr}{SDS-TWR}{Symmetrical Double Sided Two Way Ranging}
\newacronym{ds-twr}{DS-TWR}{Double Sided Two Way Ranging}
\newacronym{asym-ds-twr}{asym-DS-TWR}{Asymetric Double Sided Two Way Ranging}
\newacronym{djkm}{DJKM}{Djaja-Josko-Kolakowski-Method}
\newacronym{dpw}{DPW}{Double Pulsed Whistle}
\newacronym{mle}{MLE}{Maximum-Likelihood-Estimator}
\newacronym{dpp}{DPP}{Double Pulsed Positioning}
\newacronym{los}{LoS}{Line of Sight}
\newacronym{nlos}{NLoS}{Non Line of Sight}
\newacronym{uwb}{UWB}{Ultra Wide Band}
\newacronym{ppm}{ppm}{Parts per Million}

\usepackage{amsmath}

\usepackage{siunitx}
\usepackage{blindtext}
\usepackage{float}
\usepackage{cleveref}

\ifCLASSINFOpdf
  \usepackage[pdftex]{graphicx}
  \DeclareGraphicsExtensions{.pdf,.jpeg,.png}
\else
  \usepackage[dvips]{graphicx}
  \DeclareGraphicsExtensions{.pdf}
\fi

\ifCLASSOPTIONcompsoc
  \usepackage[caption=false,font=normalsize,labelfont=sf,textfont=sf]{subfig}
\else
  \usepackage[caption=false,font=footnotesize]{subfig}
\fi

\renewcommand\textbf{}

\makeatletter
\let\old@ps@headings\ps@headings
\let\old@ps@IEEEtitlepagestyle\ps@IEEEtitlepagestyle
\def\confheader#1{%
\def\ps@headings{%
\old@ps@headings%
\def\@oddhead{\strut\hfill#1\hfill\strut}%
\def\@evenhead{\strut\hfill#1\hfill\strut}%
}%
\def\ps@IEEEtitlepagestyle{%
\old@ps@IEEEtitlepagestyle%
\def\@oddhead{\strut\hfill#1\hfill\strut}%
\def\@evenhead{\strut\hfill#1\hfill\strut}%
}%
\ps@headings%
}
\makeatother

\confheader{2019 International Conference on Indoor Positioning and Indoor Navigation (IPIN), 30 Sept. - 3 Oct. 2019, Pisa, Italy}

\title{Clock Error Analysis of Common Time of Flight based Positioning Methods}

\IEEEoverridecommandlockouts
\IEEEpubid{\makebox[\columnwidth]
       {\hfill 978-1-7281-1788-1/19/\$31.00 ~\copyright~2019 IEEE}
\hspace{\columnsep}\makebox[\columnwidth]{ }}

\begin{document}
\twocolumn[
\centerline{\parbox[c][10cm][c]{10cm}{%
	\selectfont
	{\textbf{\LARGE{Copyright Notice}}\newline\newline
	\textcopyright 2019 IEEE. Personal use of this material is permitted. Permission from IEEE must be obtained for all
	other uses, in any current or future media, including reprinting/republishing this material for advertising
	or promotional purposes, creating new collective works, for resale or redistribution to servers or lists, or
	reuse of any copyrighted component of this work in other works.\newline\newline
	\textbf{Published in: Proceedings of the 2019 International Conference on Indoor Positioning and Indoor Navigation (IPIN), 30 Sept. - 3 Oct. 2019, Pisa, Italy}
	\newline
	\textbf{DOI: 10.1109/IPIN.2019.8911772}
}}}]

\newpage


% \author{\IEEEauthorblockN{Maximilian von Tschirschnitz}
% \IEEEauthorblockA{Technical University Munich\\
% Email: maximilian.tschirschnitz@tum.de\\
% https://orcid.org/0000-0003-4071-3630}
% \and
% \IEEEauthorblockN{Marcel Wagner}
% \IEEEauthorblockA{Intel Deutschland GmbH\\
% Germany Feldkirchen, Dornacherstr 1\\
% Email: marcel.wagner@intel.com}
% \and
% \IEEEauthorblockN{Georg Carle}
% \IEEEauthorblockA{Technical University Munich\\
% Email: carle@net.in.tum.de}
% \and
% \IEEEauthorblockN{Marc-Oliver Pahl}
% \IEEEauthorblockA{Technical University Munich\\
% Email: pahl@net.in.tum.de}}
\author{\IEEEauthorblockN{Maximilian von Tschirschnitz\IEEEauthorrefmark{1},
Marcel Wagner\IEEEauthorrefmark{2}, Marc-Oliver Pahl\IEEEauthorrefmark{3} and
Georg Carle\IEEEauthorrefmark{4}}

\IEEEauthorblockA{Technical University Munich,
Intel Deutschland GmbH\\
Email: \IEEEauthorrefmark{1}maximilian.tschirschnitz@tum.de,
\IEEEauthorrefmark{2}marcel.wagner@intel.com,
\IEEEauthorrefmark{3}pahl@s2o.net.in.tum.de,
\IEEEauthorrefmark{4}carle@net.in.tum.de}}


%Telephone: (800) 555--1212, Fax: (888) 555--1212}


\setlength{\topskip}{5pt}

\maketitle
\vspace*{-0.8cm}

\begin{abstract}
Today, many applications such as production or rescue settings rely on highly accurate
entity positioning.
Advanced Time of Flight (ToF) based positioning methods provide high-accuracy localization
of entities.
A key challenge for ToF based positioning is to synchronize the clocks between the participating
entities.\\
This paper summarizes and analyzes ToA and TDoA methods with respect to clock error robustness. The focus is on synchronization-less methods, 
i.e. methods which reduce the infrastructure requirement significantly. 
We introduce a unified notation to survey and compare the relevant work from literature. Then we apply a clock error model and compute worst case location-accuracy errors. 
Our analysis reveals 
a superior error robustness against clock errors for so called Double-Pulse methods when applied to radio based ToF positioning. 
\end{abstract}


% \fi
\IEEEpeerreviewmaketitle

\section{Introduction}
The ability to locate a device's position is highly valuable in our modern and connected world.
A considerable amount of research has therefore been conducted on that field, most recently in particular in the area
of \gls{tof} based positioning. In a typical \gls{tof} setup, radio signals are used to estimate distances between so called \emph{Anchor} and \emph{Tag} nodes. Anchors have known positions and

Tags are to be located. There are two main approaches in \gls{tof}, which lead to different solving algorithms. The first one is \gls{toa} 
which measures the sending and receiving time of a signal and is using these values to calculate a distance between two devices. 
For instance, in a 2D space the distances of one Tag to three Anchors are required to locate the Tag with multilateration algorithms \cite{murphy1995determination}. 
In \gls{toa} the time of signal transmission as well as time of arrival has to be measured. In this basic version, all Anchors and Tags need to have synchronized 
clocks to perform the measurement.

The second approach is \gls{tdoa} which measures the difference of signal arrivals. For instance, in a 2D space a signal sent from a Tag which is 
arriving at three different Anchors can be located by hyperbolic solver algorithms \cite{chan_solver}. \gls{tdoa} approaches do not require the time of signal transmission.
Only the differences of the arrival times of the signal at the Anchors need to be known. Unlike the \gls{toa} case, this requires the Anchors to be synchronized among each other.
No synchronization between Anchors and Tag is needed.  

To have synchronized clocks is a considerable infrastructure requirement and, therefore, alternative methods to do synchronization-less \gls{toa} and \gls{tdoa} 
have been developed \cite{hach_sds_twr, ds_sds_twr, whistle, dpw, anti_whistle}.
Moreover, the fact that some of these methods work with radio waves, some with ultrasound is posing different requirements on the  tolerable clock accuracy.  
The goal of this work is to analyze the relevant \gls{toa} and \gls{tdoa} methods from the field of synchronization-less \gls{tof}.
% In order to establish a founding understanding of the topic and to point out relationships between the methods, which
% may lead to new accomplishments in the field.
To achieve this, an error model is defined and applied to the approaches in order compare them against each other and to point out commonalities and differences.

The rest of the paper is organized as follows. First we will introduce the error model (Section \ref{sct:unified_error_model}). 
Then both, \gls{toa} and \gls{tdoa} methods will be presented in Section \ref{sct:toa} and \ref{sct:tdoa} respectively. Finally, in Section \ref{sct:comparision}, 
a comparison and conclusion are given.

\section{Unified Error Model}\label{sct:unified_error_model}
	Our error model is inspired by the works of \citet{alternative_twr} and \citet{dpw}.
	We adopt and optimize their underlying principle to make it applicable to the methods 
	we assess in this paper.

	\subsection{Clock Drift and Synchronization}\label{sct:clock_drift_and_sync}
		Every clock contains imperfections, making it run at an inconstant rate.
		No existing clock is able to keep the perfect time \cite{kamas1979time}.
		Since \gls{tof}-based methods rely on the measurements of signals propagating with high speeds such as the speed of sound or the speed of light ($c$),
		clock imperfections make the runtime measurements inaccurate.

		%Another issue with the so-called clock drift is the resulting continuous de-synchronization of distributed entities with originally synchronized clocks.
		
		To reach a better understanding of clock errors, their deviation is typically described in the \emph{parts-per} notation which is in the following  
    in the order of magnitude of millions, hence \gls{ppm} \cite{ppm}.
		The parts-per notation specify the ratio of how many clock ticks the current clock-value is expected to deviate from the real time.
		As notation for the rest of this paper we introduce the  symbol of $\epsilon_A$ to describe the clock error of device or oscillator $A$.
		%We use the symbol $x$ to describe the error factor introduced by the clock drift.
		For quartz oscillators a typical value for $\epsilon$ is $\pm20$ \gls{ppm} as defined in IEEE802.15.4 \cite{ieee_twr}.

		The current time $\hat{t}$ of device $A$ can therefore be expressed as:
		$$\hat{t} := t\label{eqn:error_modulation} \cdot (1 + \epsilon_A)$$
		%\textbf{TODO: Is this X thee same as above? If so why is the error factor used as index??? If not use another letter!}
		$t$ is the ideal, error-free clock counter value and
		$\hat{t}$ is the error-affected counterpart.
		
		Since in this paper we focus on \gls{tof}, which implies measurement of propagation of radio waves,
		clock counter precision needs to be below nanoseconds.
		For instance, $1$ns corresponds to $\approx 0.3$m of signal propagating with the speed of light.
		
		A typical assumption in the field of \gls{tof} is that the drift of a specific clock is constant to allow measurements with the same clock error $\epsilon$ for some time \cite{alternative_twr}. 
		%This assumption is in line with practical experiments we did with physical hardware.

	\subsection{Measurement Errors and Multipath}
		The accuracy of time of flight measurements can be affected by additional error factors.
		 The two most relevant factors are \gls{nlos} errors and multipath-propagation effects.

		\paragraph{Multipath-Propagation Effect}
		The multipath-effect describes the fact that electromagnetic signals can reach a receiver on multiple paths.
		They can be reflected by walls or other obstacles for instance.
		As a consequence, a measured signal propagation time is not always that of the shortest possible path between the emitter and the receiver.
		However, for accurate distance measurements, the identification of the direct signal path, the so called \emph{primary signal} is essential.
		
		\paragraph{\gls{nlos} Error}
		%Measurement errors are randomly occurring errors.
		%They can have different causes such as
		%\textbf{TODO: namme 1-2 causes.} 
		%measurement errors influence the timestamping process
		%\textbf{TODO: introduce timestamping!}
		%in an unpredictable way.
		%As they are random, they cannot be eliminated through calibration.
		\gls{nlos} errors occur if the direct path between a sender and a receiver is blocked by a material with different propagation properties than for example air.
		The signal traveling on the direct path between sender and receiver goes through the obstacle before reaching the antenna of the receiver.
		The obstacle changes the propagation speed.
    
		As a consequence of having such material changes in real-world settings,  it is impossible to make accurate assumptions %(see \ref{sct:dev_pos_and_tof})
		about the signals traveling speed between the sender and the receiver. 
		Moreover, \gls{nlos} scenarios make the identification of the primary signal under multipath-propagation harder. 

		Though the previously described problems have high relevance, 
		we assume in the following \gls{los} scenarios where the measurement errors can be neglected.
		%We also assume that to work with specialized hardware such as Decawave's DW1000 \cite{dw_manual} chip.
		%It is able to identify the direct path from a set of multipath-propagated signals
		%in \gls{los} scenarios.
		%When not explicitly noted, we do not consider multipath-propagation and measurement errors in the remainder of the paper.

		A more detailed description and analysis of those effects can be found in \cite{dw_multipath_nlos}.

\section{Time of Arrival}\label{sct:toa}
\label{TOA}
	Having defined this error-model we can now apply it to the relevant \gls{tof} methods from the state of the art.
	We will group the analyzed methods into \gls{toa} and \gls{tdoa}, beginning with the former one.
	%Both, \gls{toa} and \gls{tdoa} methods generate output which can be used as input to so called multilateration methods in order to determine
	%the final position of a device (called Tag) most often under knowledge about the position of certain infrastracture devices (called Anchors).

	\subsection{Simple Time-of-Arrival}\label{sct:simple_toa}
\begin{figure}
			\centering
			\includegraphics{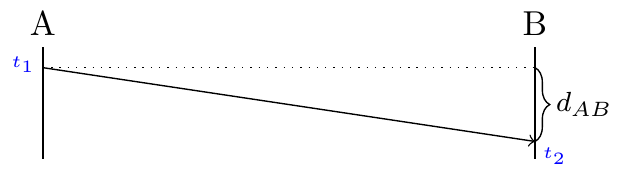}
			\caption{Simple \gls{toa} scheme}
			\label{fig:simple_toa}
		\end{figure}

		The simplest \gls{tof} setup imaginable would be the one of simple Time-of-Arrival (simple-\gls{toa}).
		Simple-\gls{toa} consists of two devices $A$ and $B$ between which one signal is exchanged (compare \textbf{Fig. \ref{fig:simple_toa}}).
		This signal sent by $A$ is timestamped (using $A$s clock) at the moment of transmission as well as when it is received at $B$ (using $B$s clock).
		If we now consider for a moment the synchronization between the devices clocks to be perfect, the difference between $A$ and $B$s timestamps
		would resemble the exact \gls{tof} between both.
		\begin{IEEEeqnarray}{rCl}
			d_{AB} = t_2 - t_1 \label{eqn:simple_toa_eqn}
		\end{IEEEeqnarray}
		Then the distance between both could be calculated as:
		\begin{IEEEeqnarray}{rCl}
			\overline{AB} = c \cdot d_{AB} \label{eqn:simple_toa_eqn_dist}
		\end{IEEEeqnarray}
		Problems arise if we now reconsider the calculation under the assumption that synchronization is not provided and also when considering the presence of clock drift effects (like described in Section \ref{sct:clock_drift_and_sync}).
		For example, even if we assume to have both clocks to be synchronized at one distinct moment ($t = 0$),
		the error would then accumulate from that moment on.
		We can calculate that worst-case error in the final distance estimate using our clock error model (Section \ref{sct:unified_error_model}).

		Like in \eqref{eqn:error_modulation} we define a clock error model:
		\begin{align*}
			\hat{t}_1				&:= t_1 \cdot (1 + \epsilon_A) 															\\
			\hat{t}_2				&:= t_2 \cdot (1 + \epsilon_B)															\\
		\end{align*}
		Based on these erroneous timestamps and using \eqref{eqn:simple_toa_eqn} and \eqref{eqn:simple_toa_eqn_dist} we then define the erroneous \gls{toa} ($d_{AB}$) and distance value ($\overline{AB}$) as:
		\begin{align*}
			\hat{d}_{AB}			&:= \hat{t}_2 - \hat{t}_1															\\
			\hat{\overline{AB}} 	&:= c \cdot \hat{d}_{AB}																	\\
		\end{align*}
		Consequently, we define the difference between the erroneous and error-free value as the error.
		\begin{align*}
			\tilde{\overline{AB}}	&:= \hat{\overline{AB}} - \overline{AB} \\
											&= c \cdot (t_2 (1 + \epsilon_B) - t_1 (1 + \epsilon_A) - t_2 + t_1)	\\
											&= c \cdot \epsilon_B t_2 - c \cdot \epsilon_A t_1
		\end{align*}
		Moreover, using the fact that $t_2 = t_1 + d_{AB}$ we can substitute $t_2$:
		\begin{align*}
			&= c \cdot \epsilon_B (t_1 + d_{AB}) - c \cdot \epsilon_A t_1 \\
			&= c \cdot \epsilon_B t_1 - c \cdot \epsilon_A t_1 + c \cdot \epsilon_B d_{AB}\\
		\end{align*}
		And since $t_1 \gg d_{AB}$ ($d_{AB}$ is in order of magnitude of nanoseconds) in almost all realistic scenarios, we get the approximation:
		\begin{align*}
			&\approx c \cdot t_1(\epsilon_B - \epsilon_A)
		\end{align*}
		When assuming the default clock drift of $\pm20$\gls{ppm} the worst value $(\epsilon_B - \epsilon_A)$ can take on is $\frac{40}{1000000}$.
		So the worst-case error introduced can be estimated as:
		$$t_1 \cdot \num{3e8}\frac{\text{m}}{\text{s}} \cdot \frac{40}{1000000} = t_1 \cdot \num{12e3}m$$
		Therefore we expect an worst case error of $\num{12e3}$ meters for every second elapsed on $t_1$ on from the moment of the last synchronization until the moment the measurement took place. 
		This effect would without question, render such a system for almost all applications unusable.
		%Please also note that the assumption of a synchronization at some point of operation already requires considerable technical effort in practice already.

	\subsection{Two-Way-Ranging}\label{sct:twr}
\begin{figure}
			\centering
			\includegraphics{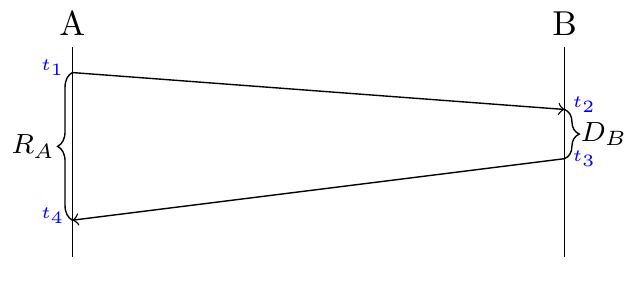}
			\caption{Simple \gls{twr} scheme}
			\label{fig:simple_twr}
		\end{figure}

		A first simple solution to compensate the missing synchronization between devices is the concept of \gls{twr} as determined in IEEE802.15.4 described in \cite{ieee_ss_twr}.
		\gls{twr} systems include two devices $A$ and $B$, that are communicating bi-directional to measure a round-trip-time.
		This concept is simply described by $A$ sending a message at $t_1$ to $B$ (received at $t_2$) which gets then acknowledged after a fixed and known delay
		at time $t_3$ (sent from $B$).
		That acknowledgment arrives back at device $A$ at $t_4$.
		According to \cite{ieee_ss_twr}, $A$ can now calculate the \gls{tof} as follows:
		\begin{IEEEeqnarray}{rCl}
			d_{AB} = \frac{1}{2} \cdot (R_A - D_B) \label{eqn:twr_eqn} 
		\end{IEEEeqnarray}
		with
		$R_A := t_4 - t_1$ and $D_B := t_3 - t_2$ (comp. \textbf{Fig. \ref{fig:simple_twr}})

		This method is expected to be better conditioned since we do not rely on the clocks being synchronized at the beginning of the measurement.
		Therefore the error should not accumulate in the same way as in simple \gls{toa} (Section \ref{sct:simple_toa}).
		This intuition is now substantiated.
		Similar as in Section \eqref{eqn:error_modulation} we define an erroneous model for the here relevant values:
		\begin{align}
			\hat{R}_A				&:= R_A \cdot (1 + \epsilon_A) \label{eqn:simple_twr_ra} \\
			\hat{D}_B				&:= D_B \cdot (1 + \epsilon_B) \label{eqn:simple_twr_db}
		\end{align}
		And by using \eqref{eqn:twr_eqn} we can define:
		\begin{align*}
			\hat{d}_{AB}			&:= \frac{1}{2} (\hat{R}_A - \hat{D}_B)\\
										&=  \frac{1}{2} \Big[(\hat{t}_4 - \hat{t}_1) - (\hat{t}_3 - \hat{t}_2)\Big]\\
			\hat{\overline{AB}} 	&:= c \cdot \hat{d}_{AB}
		\end{align*}
		Again the difference between erroneous and error-free value states the error.
		\begin{align*}
			\tilde{\overline{AB}} &= \hat{\overline{AB}} - \overline{AB} \\
										 &= c \cdot \frac{1}{2} (\hat{R}_A - R_A - \hat{D}_B + D_B) \\
										 &= c \cdot \frac{1}{2} (\epsilon_A R_A  - \epsilon_B D_B) \text{ by using \eqref{eqn:simple_twr_ra} and \eqref{eqn:simple_twr_db}}
		\end{align*}
		Using the fact that $R_A = D_B + 2d_{AB}$ it follows:
		\begin{align*}
										 &= c \cdot \frac{1}{2} \Big[\epsilon_A(D_B + 2d_{AB}) - \epsilon_B D_B\Big] \\
										 &= c \cdot \frac{1}{2} \Big[(\epsilon_A - \epsilon_B) D_B + \epsilon_A 2d_{AB}\Big]
		\end{align*}
		And since $D_B \gg d_{AB}$ in realistic scenarios, we get the approximation:
		\begin{align*}
										 &\approx c \cdot \frac{1}{2} (\epsilon_A - \epsilon_B) \cdot D_B
		\end{align*}
		With clock-drift-error the worst value $(\epsilon_A - \epsilon_B)$ can take on is $\frac{40}{1000000}$, so the worst case error introduced
		can be calculated as:
		$$\num{3e8}\frac{\text{m}}{\text{s}} \cdot \frac{1}{2} \cdot \frac{40}{1000000} \cdot D_B = \num{6e3}\text{m} \cdot D_B$$
		% We approximated $D_B $ with $R_B$, this is valid since $d_{AB}$ is expected to be orders of magnitudes smaller then both $R_A$ and $D_B$.
		We observe that the clock error is only affecting the measurement during $D_B$ (on both clocks) and does
		not accumulate over the whole operation time.
		Nevertheless, we still consider the system too error prone in most cases since even a $D_B$ of one millisecond causes already six meters of error.
		A similar analysis is shown in \cite[D1.3.1]{ieee_twr}.

	\subsection{Double-Sided Two-Way-Ranging}\label{sct:ds_twr}
		\subsubsection{Concept}
\begin{figure}
			\centering
			\includegraphics{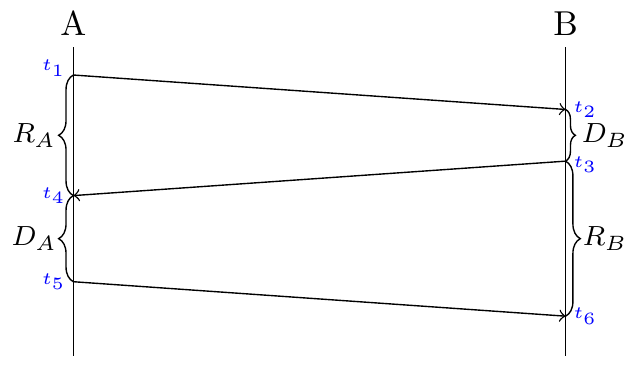}
			\caption{Double-Sided \gls{twr} scheme}
			\label{fig:ds_twr}
		\end{figure}

		To reduce the influence of drift-error in \gls{twr}, a new method called \gls{ds-twr} is proposed.
		This method is described and defined in the IEEE 802.15.4 standard \cite[D1.3.2]{ieee_twr}.
		%as the result of the works of \citet{hach_sds_twr}.
		They suggested adding a third message to the scheme as depicted in \textbf{Fig. \ref{fig:ds_twr}}.
		Using this method the $d_{AB}$ respectively $\overline{AB}$ can (in the ideal case) be calculated as follows from \citet{hach_sds_twr}:
		\begin{IEEEeqnarray}{rCl}
		d_{AB} = \frac{1}{4}(R_A - D_B + R_B - D_A) \label{eqn:ds_twr_eqn}
		\end{IEEEeqnarray}
		\begin{IEEEeqnarray}{rCl}
		\overline{AB} = c \cdot d_{AB} \label{eqn:ds_twr_eqn_dist}
		\end{IEEEeqnarray}
		Similar to the previous method we can also calculate the error margin.
		First, we create a clock error model for this method like in \eqref{eqn:error_modulation}:
		\begin{equation}
			\label{eqn:ds-twr_error_model}
			\begin{array}{ccl}
				\hat{R}_A := R_A \cdot (1 + \epsilon_A) \\
				\hat{D}_A := D_A \cdot (1 + \epsilon_A) \\
				\hat{R}_B := R_B \cdot (1 + \epsilon_B) \\
				\hat{D}_B := D_B \cdot (1 + \epsilon_B)
			\end{array}
		\end{equation}
		And by using \eqref{eqn:ds_twr_eqn} and \eqref{eqn:ds_twr_eqn_dist} we derive:
		\begin{align*}
			\hat{d}_{AB}			&:= \frac{1}{4}(\hat{R}_A - \hat{D}_B + \hat{R}_B - \hat{D}_A) \\
			\hat{\overline{AB}} 	&:= c \cdot \hat{d}_{AB}
		\end{align*}
		Again the difference between erroneous and error-free value states the error.
		\begin{align*}
			\tilde{\overline{AB}} &= \hat{\overline{AB}} - \overline{AB} \\
										 &= c \cdot \frac{1}{4}\Big[(R_A - D_A) \cdot \epsilon_A + (R_B - D_B) \cdot \epsilon_B\Big]
		\end{align*}
		We then add and subtract to the inner brackets $D_B$ respectively $D_A$:
		\begin{align*}
										 = &\;c \cdot \frac{1}{4}\Big[(R_A - D_B + D_B - D_A) \cdot \epsilon_A\\
										 &+ (R_B - D_A + D_A - D_B) \cdot \epsilon_B\Big] \\
										 = &\;c \cdot \frac{1}{4}\Big[(R_A - D_B) \cdot \epsilon_A + (D_B - D_A) \cdot \epsilon_A\\
										 &+ (R_B - D_A) \cdot \epsilon_B + (D_A - D_B) \cdot \epsilon_B\Big]
		\end{align*}
			And since $2d_{AB} = R_A - D_B$ as well as $2d_{AB} = R_B - D_A$ we can substitute those:
		\begin{align*}
										 &= c \cdot \frac{1}{4}\Big[2d_{AB} \cdot (\epsilon_A + \epsilon_B) + (D_B - D_A) \cdot (\epsilon_A - \epsilon_B)\Big] \\
										 &= c \cdot \frac{1}{4}\Big[(D_B - D_A) \cdot (\epsilon_A - \epsilon_B)\Big] + c \cdot \frac{1}{4}\Big[2d_{AB} \cdot (\epsilon_A + \epsilon_B)\Big] 
		\end{align*}
		$d_{AB}$ is in order of magnitude of nanoseconds and multiplied with a $\approx20$\gls{ppm} value, so it follows that the second term is in sub-picosecond order of magnitude.
		Even under multiplication with the large value of $c$, this would leave that part of the term as not significant for our accuracy requirements.
		Therefore only the first part of the term is relevant to us:
		\begin{align*}
										 &\approx c \cdot \frac{1}{4}(D_B - D_A) \cdot (\epsilon_A - \epsilon_B)
		\end{align*}

		Again assuming $\pm20$\gls{ppm} as drift error (comp. Section \ref{sct:clock_drift_and_sync}), the worst case value can then be calculated:
		$$\num{3e8}\frac{\text{m}}{\text{s}} \cdot \frac{1}{4}(D_B - D_A) \cdot \frac{40}{1000000} = \num{3e3}\text{m} \cdot (D_B - D_A)$$
		Where $(\epsilon_A - \epsilon_B)$ can take on the value $\frac{40}{1000000}$ in the worst case.
		So as long as the difference $|D_A - D_B|$ is relatively small, we can assume a very low error.

		In the following, we denote this method as \gls{sds-twr}.
		With this method, it is at least in theory, possible to eliminate the drift error completely by selecting $D_A$ and $D_B$ to be identical.
		However, this is, in reality not possible (e.g. due to inaccuracy of scheduling sending time).
		To remove this requirement, the asymmetric approach was developed.

		\subsubsection{Asymmetric Formula}\label{sct:asym_ds_twr}
			Keeping the difference $|D_A - D_B|$ small is not always possible or desired (for instance through technical limitations of the measuring/processing device).
			To address those restrictions, \citet{alternative_twr} developed a different (asymmetric) formula for the same scheme:
			\begin{align}
				d_{AB} :&= \frac{R_A R_B - D_A D_B}{2(R_A + D_A)} \label{eqn:asym_ds_twr_eqn_a} \\
						 &= \frac{R_A R_B - D_A D_B}{2(R_B + D_B)} \label{eqn:asym_ds_twr_eqn_b} \\
				\overline{AB} &:= c \cdot d_{AB} \label{eqn:asym_ds_twr_eqn_dist}
			\end{align}
			$\Rightarrow$ Using $D_B + R_B = R_A + D_A$ (comp. Fig. \ref{fig:ds_twr})
			\begin{IEEEeqnarray}{rCl}
				d_{AB} = \frac{R_A R_B - D_A D_B}{R_A + R_B + D_A + D_B} \label{eqn:asym_ds_twr_eqn}
			\end{IEEEeqnarray}
			When this adjusted formula is applied, we refer to the method as \gls{asym-ds-twr}.
			The formula is in the following proven to be more resilient to differences between $D_A$ and $D_B$.

			% First we define an clock error model like in \eqref{eqn:error_modulation}:
			% \begin{align*}
			% 	\hat{R}_A				&:= R_A \cdot (1 + \epsilon_A) \\
			% 	\hat{D}_A				&:= D_A \cdot (1 + \epsilon_A) \\
			% 	\hat{R}_B				&:= R_B \cdot (1 + \epsilon_B) \\
			% 	\hat{D}_B				&:= D_B \cdot (1 + \epsilon_B)
			% \end{align*}
			Using the same clock error model as for the symmetric solution \eqref{eqn:ds-twr_error_model} and applying it to \eqref{eqn:asym_ds_twr_eqn_a} we define:
			\begin{align*}
				\hat{d}_{AB}	&:= \frac{\hat{R}_A \hat{R}_B - \hat{D}_A \hat{D}_B}{2(\hat{R}_A + \hat{D}_A)} \\
									&= \frac{(1 + \epsilon_A)(1 + \epsilon_B)}{(1 + \epsilon_A)} \cdot \frac{R_A R_B - D_A D_B}{R_A + D_A}\\
			\end{align*}
			And equivalently using \eqref{eqn:asym_ds_twr_eqn_b} we derive:
			\begin{align*}
											&= \frac{\hat{R}_A \hat{R}_B - \hat{D}_A \hat{D}_B}{2(\hat{R}_B + \hat{D}_B)} \\
											&= \frac{(1 + \epsilon_A)(1 + \epsilon_B)}{(1 + \epsilon_B)} \cdot \frac{R_A R_B - D_A D_B}{R_B + D_B}
			\end{align*}
			With \eqref{eqn:asym_ds_twr_eqn_dist} it follows the definition:
			\begin{IEEEeqnarray}{rCl}
				\hat{\overline{AB}}	:= c \cdot \hat{d}_{AB}
			\end{IEEEeqnarray}
			The difference between erroneous and error-free value states the absolute error.
			\begin{align*}
				\tilde{\overline{AB}}&= \hat{\overline{AB}} - \overline{AB} \\
											&= c \cdot \epsilon_Ad_{AB} \text{ using \eqref{eqn:asym_ds_twr_eqn_a}}\\
				\text{and}&\\
											&= c \cdot \epsilon_Bd_{AB} \text{ using \eqref{eqn:asym_ds_twr_eqn_b}}\\
			\end{align*}
			Like in \eqref{eqn:asym_ds_twr_eqn} we can combine those:
			\begin{align*}
				\hat{\overline{AB}}	&= c \cdot \Big[\frac{1}{2}(\epsilon_A + \epsilon_B) \cdot d_{AB}\Big] \\
											&= \frac{\epsilon_A + \epsilon_B}{2} \cdot \overline{AB}
			\end{align*}
			So in the worst case we can assume to have an error of $\frac{20}{1000000} \cdot \overline{AB}$ which is non-significant in almost all cases.
			This is achieved without having similar $D_A$ and $D_B$.
			It is important to note that we assumed a constant clock drift during the time of measurement in our clock-model (\ref{sct:clock_drift_and_sync}) . 

\section{Time Difference of Arrival}\label{sct:tdoa}
\label{TDOA}
	Beside the \gls{toa} methods described there is the group of \gls{tdoa}-based methods.

	In contrast to \gls{toa}, \gls{tdoa} methods have less strict device capability requirements.
	For example, some of the following methods allow to position Tags which are not actively responding in reaction to the Anchors' signals, while some methods
	allow the Anchor network to stay silent (no transmission) during positioning.
	A few beneficial features resulting from that are for instance the possibility to increase the number of Nodes
	that are simultaneously positioned or the ability to position devices without their knowledge or devices which are incapable of measuring time. 

	The difference between \gls{toa} and \gls{tdoa} becomes clear when looking at an example for a simplified version of a \gls{tdoa} method.
	
	\subsection{Simple TDoA}\label{sct:simple_tdoa}
\begin{figure}
			\centering
			\includegraphics{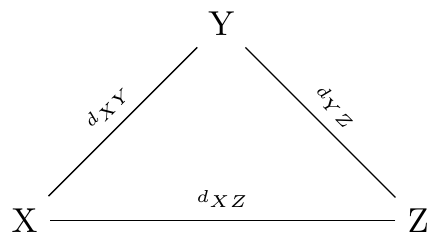}
			\caption{TDoA value graphically}
			\label{fig:tdoa_triangle}
		\end{figure}

		Simple-\gls{tdoa} is a method in which we do not measure the \gls{tof} between two devices.
		Instead, we measure the difference in distances/time between each of two devices $X$ and $Z$ to one device $Y$ (comp. \textbf{Fig. \ref{fig:tdoa_triangle}}).
		This value is called a Time-Difference-of-Arrival- or short \gls{tdoa}-value and is in this document denoted as $T_{XZ}^Y$.
		% Again as with \gls{toa}, multiple of such \gls{tdoa}-values (between a Tag device and multiple Anchor devices) ultimately allow
		% us the determination of the Tags position through multilateration methods like in \ref{chan_solver}.

\begin{figure}
			\centering
			\includegraphics{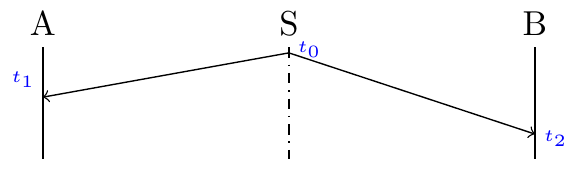}
			\caption{Simple TDoA}
			\label{fig:simple_tdoa}
		\end{figure}

		The mathematical definition of this \gls{tdoa} value is:
		$$T_{XZ}^Y := d_{YZ} - d_{YX}$$
		The simplest setup in which we would effectively gather that value would consist of three perfectly synchronized and drift-free devices $A, B, S$
		from which $S$ transmits a message at a time $t_1$, and the signal arrives at $A$ at time $t_1$, and at $B$ at time $t_2$ as depicted in \textbf{Fig. \ref{fig:simple_tdoa}}
		We can then calculate:
		\begin{IEEEeqnarray}{rCl}
			T_{AB}^S = t_2 - t_1  \label{eqn:def_simple_tdoa}
		\end{IEEEeqnarray}
		% Under knowledge of the positions of $A$ and $B$ as well as the signals speed $v$ this value can be used to describe a hyperboloid in the measurement space (comp. \cite{geometric_tdoa_solution}).
		% %on which in one point$S$ would have to reside on.
		% The transmitting device $S$ is here interpreted as Tag while $A$ and $B$ are operating as Anchors.
		% % The replying anchor device, here $B$, is characterized more specifically as \emph{Mirror}.
		% Multiple such measurements on additional Anchor devices, in relation to the same Tag, would
		% provide us with multiple such hyperboloids.
		% The common intersection point of those yields us then the true point position of $S$.
		% To solve this hyperboloid-intersection problem there is a variety of approaches which we examine in Section \ref{sct:multilateration}.
		The transmitting device $S$ is here interpreted as Tag while $A$ and $B$ are operating as Anchors.
		Multiple such measurements on additional Anchor devices (with known positions), in relation to the same Tag, would allow to apply hyperbolic solver algorithms 
    like \citet{chan_solver} to determine the Tags position. 

		When calculating the error of this setup we have to assume that since the last perfect synchronization ($t = 0$), drift effects
		have been accumulated by the individual clocks.

		As with \gls{toa} methods previously we now primarily define our erroneous model like in \eqref{eqn:error_modulation}:
		\begin{align*}
			\hat{t}_1 &:= t_1 \cdot (1 + \epsilon_A) \\
			\hat{t}_2 &:= t_2 \cdot (1 + \epsilon_B)
		\end{align*}
		And using \eqref{eqn:def_simple_tdoa}: 
		\begin{align*}
			\hat{T}_{AB}^S &:= \hat{t}_2 - \hat{t}_1 \\ 
		\end{align*}
		Again the difference between erroneous and error-free value states the error.
		\begin{align*}
			\tilde{T}_{AB}^S &= \hat{T}_{AB}^S - T_{AB}^S \\
					&= \epsilon_Bt_2 - \epsilon_At_1 \\
		\end{align*}
		Using the fact that $t_2 = t_1 + T_{AB}^S$ we can write
		\begin{align*}
					&= \epsilon_B(t_1 + T_{AB}^S) - \epsilon_At_1 \\
					&= t_1(\epsilon_B - \epsilon_A) + \epsilon_{B}T_{AB}^S\\
		\end{align*}
		And since $T_{AB}^S$ is in order of magnitude of nanoseconds and is multiplied with a $\approx20$\gls{ppm} value the second part of the term is in sub-picosecond order of magnitude.
		Therefore only the first part of the term appears to be relevant, and we can approximate it with:
		\begin{IEEEeqnarray}{rCl}
					t_1 \cdot (\epsilon_B - \epsilon_A)
		\end{IEEEeqnarray}
		That means that we have to assume a worst-case error of $\approx \frac{40}{1000000} \cdot t_1$.
	 	%  This error would quickly (seconds) amount to thousend of meters of error and would make any further processing (for instance using a \gls{tdoa} solver) unpractical. 
		Which makes the method already after a short time of operation unusable for realistic scenarios.
		% Furthermore we need to keep in mind that $t_1$ is counting on magnitudes of nanosecond according to our clock model (\ref{sct:clock_drift_and_sync}).
		% So after one second of operation time the error already becomes very large and the method is not realistically useable for positioning tasks any more.

	\subsection{Whistle}\label{sct:whistle}
\begin{figure}
			\centering
			\includegraphics{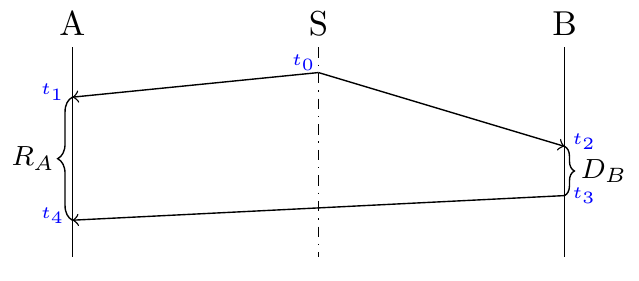}
			\caption{Whistle}
			\label{fig:whistle}
		\end{figure}

		To resolve that issue of fast degrading synchronization.
		\citet{whistle} proposed a method called \textbf{Whistle}.
		Whistle tries to reduce the timespan in which the synchronization drift can occur.
		That is achieved by adding another signal into the simple \gls{tdoa} scheme as shown in \textbf{Fig. \ref{fig:whistle}}

		This additional message exchange between $A$ and $B$ can be interpreted as a form of ``resynchronization'' between both devices.
		According to \citet{whistle} the \gls{tdoa} value between those devices can now be calculated by:
		\begin{align}
		T_{AB}^S &:= d_{AB} - (t_4 - t_1) + (t_3 - t_2) \nonumber \\
					&= d_{AB} - R_A + D_B\label{eqn:whistle_tdoa}
		\end{align}
		The Anchor that is replying (here $B$) is called \textbf{Mirror}.
		Note that we require that the distance between the Mirror and the other Anchor devices is known.

		Now we calculate the error resulting from this improved protocol.
		First, we define the erroneous model like in \eqref{eqn:error_modulation}:
		\begin{align*}
			\hat{R}_A &:= R_A \cdot (1 + \epsilon_A) \\
			\hat{D}_B &:= D_B \cdot (1 + \epsilon_B)
		\end{align*}
		Then using \eqref{eqn:whistle_tdoa}: 
		\begin{align*}
			\hat{T}_{AB}^S &:= d_{AB} - \hat{R}_A + \hat{D}_B \\
		\end{align*}
		Again the difference between erroneous and error-free value states the error.
		\begin{align*}
			\tilde{T}_{AB}^S &:= \hat{T}_{AB}^S - T_{AB}^S \\ 
					&= \epsilon_BD_B - \epsilon_AR_A
		\end{align*}
		And using the fact that $R_A = D_B + d_{AB} + T_{AB}^S$ it follows:
		\begin{align*}
					&= \epsilon_BD_B - \epsilon_A(D_B + d_{AB} + T_{AB}^S) \\ 
					&= \epsilon_A(d_{AB} + T_{AB}^S) + D_B(\epsilon_B - \epsilon_A)
		\end{align*}
		$d_{AB}$ and $T_{AB}^S$ are in order of magnitude of nanoseconds.
		And since those are multiplied with a $\approx20$\gls{ppm} value in the first part of the term, that part of the term is in sub-picosecond order of magnitude.
		Therefore only the second part of the term is relevant.
		\begin{IEEEeqnarray}{rCl}
					\approx D_B(\epsilon_B - \epsilon_A) \label{eqn:whistle_error_estimate}
		\end{IEEEeqnarray}

		Assuming the same realistic worst case drift values as previously we end up with a worst-case error estimate of $D_B \cdot \frac{40}{1000000}$.
		This is an improvement compared to the simple \gls{tdoa} method since the error is not dependent on synchronized clocks and is therefore
		also not accumulating.
		On the other hand, a typical value for $D_B$ is one millisecond which leads already to $40$ nanoseconds $\approx 12$ meters of error.
		This approach is therefore not practically usable with radio platforms that operate with industry standard clocks.

	\subsection{Djaja-Josko-Kolakowski Method}\label{sct:djkm}
		A method combining the benefits of \gls{sds-twr} and Whistle was proposed by \citet{anti_whistle}, and therefore
		we refer to it as \textbf{\gls{djkm}}.

\begin{figure}
			\centering
			\includegraphics{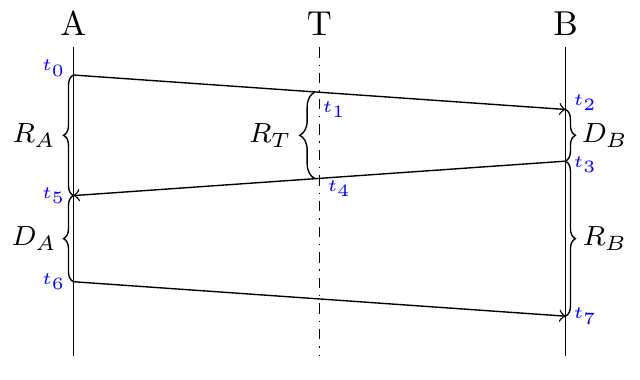}
			\caption{Djaja-Josko-Kolakowski Method}
			\label{fig:anti_whistle}
		\end{figure}

		The essential idea of \gls{djkm} is to execute an optimized version of \gls{sds-twr} in a group of Anchor devices.
		While doing so, it is possible for eavesdropping Tags to collect \gls{tdoa} information to those Anchors.
		To simplify the explanation of the method the sequence diagram \textbf{Fig. \ref{fig:anti_whistle}} was reduced to two Anchors
		(one of them could be interpreted as ``acting Mirror'') and a single Tag.
		For details about the interaction between more than two Anchors we refer to the original paper \cite{anti_whistle}. 

		The authors derived the following formula to calculate the \gls{tdoa} values:
		\begin{align}
			T_{AB}^T &= (t_4 - t_1) - (t_3 - t_2) - d_{AB} \nonumber\\
						&= R_T - D_B - d_{AB}\label{eqn:djkm_eqn}
		\end{align}
		It is important to note that Tag $T$ is only in receiving mode which is an essential functional difference to Whistle where the Tag $S$ is at all
		times in transmit mode only.
		Moreover, when comparing \gls{djkm}s \gls{tdoa} calculation with the one from Whistle \eqref{eqn:whistle_tdoa} and comparing the modes of operation of Anchors and Tags between both method
		\gls{djkm} appears to us like an ``inverse Whistle'' method. 
    % We therefore also refer to it as \textbf{inv-Whistle}.
		Especially in \gls{uwb} it is common that transmission operations use significantly less energy than receiving (comp. \cite[Sct 7.2]{dw_power_consumption}), so for battery driven Tag devices the Whistle approach is consuming less energy.

		Now we demonstrate that the worst-case error for \gls{djkm} is identical to Whistle.  
		Again we first define the erroneous model like in \eqref{eqn:error_modulation}:
		\begin{align*}
			\hat{R}_T &:= R_T \cdot (1 + \epsilon_T) \\
			\hat{D}_B &:= D_B \cdot (1 + \epsilon_B)
		\end{align*}
		Using \eqref{eqn:djkm_eqn} 
		\begin{align*}
			\hat{T}_{AB}^T &:=  \hat{R}_T - \hat{D}_B - d_{AB}\\
		\end{align*}
		The difference between erroneous and error-free value states the error.
		\begin{align*}
			\tilde{T}_{AB}^T &= \hat{T}_{AB}^T - T_{AB}^T \\ 
					&= \epsilon_TR_T - \epsilon_BD_B \\ 
		\end{align*}
		Using the fact that $R_T = D_B + d_{AB} + T_{AB}^T$:
		\begin{align*}
					&= \epsilon_T(D_B + d_{AB} + T_{AB}^T) - \epsilon_BD_B\\ 
					&= \epsilon_T(d_{AB} + T_{AB}^T) + (\epsilon_T - \epsilon_B)D_B \\ 
		\end{align*}
		For the same reason we described in Whistle (\ref{sct:whistle}) we can neglect the first part as not significant and so approximate the term with.
		\begin{IEEEeqnarray}{rCl}
					\approx (\epsilon_T - \epsilon_B)D_B \label{eqn:inv_whistle_error_estimate} 
		\end{IEEEeqnarray}
		When choosing worst-case epsilons, we end up with the worst case estimate $D_B \cdot \frac{40}{1000000}$.
		That is identical to the estimate for Whistle. 
		% To summarize the main differences of this method to Whistle are:
		% \begin{itemize}
		% 	\item{An \emph{Anchor} sends out the initial signal which is received by a \emph{Tag}}
		% 	\item{The role of the ``Mirror'' \emph{Anchor} is passed on in a token ring fashion}
		% 	\item{\gls{djkm} contains a \gls{sds-twr} scheme}
		% \end{itemize}

		\citet{anti_whistle} also point out that their scheme simultaneously allows for the calculation of \gls{sds-twr} values between certain Anchors in the ranging scheme.
			
	\subsection{Double-Pulsed-Whistle}\label{sct:dp_whistle}
\begin{figure}
			\centering
			\includegraphics{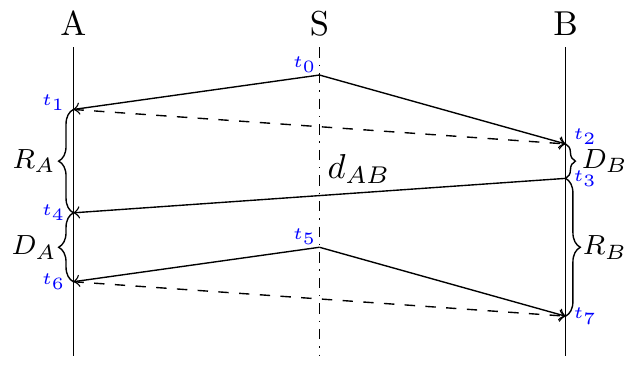}
			\caption{Double Pulsed Whistle}
			\label{fig:dpw}
		\end{figure}

		We have shown that the \gls{tdoa} values of \gls{djkm} have a similar error as Whistle.
		That means that all described methods of synchronization-free \gls{tdoa} are still prone to clock-drift errors on a scale too large for most radio-based applications.
		To circumvent these problems with \gls{tdoa} and to make the methods more applicable in practice a new approach was proposed by \citet{dpw}.
		This method called \gls{dpw} introduces a second pulse to the known Whistle scheme and uses symmetries
		to reduce the clock error significantly.
		The error reduction uses similar effects like the methods of DS-\gls{twr}.
		% And there are similiarities with \gls{djkm} on which we elaborate in section \ref{sct:djkm_protocol}
		The transmission scheme is displayed in \textbf{Fig. \ref{fig:dpw}}.

	\begin{table*}
		\center
		\caption{Comparison between the separate methods}
		\begin{tabular}{|c|c|c|c|}  
			\cline{2-4}
			\multicolumn{1}{c|}{} & Clock-Drift induced Error & Anchor Nodes & Tag Nodes  \\
			\hline
			Simple \gls{toa} ((\ref{sct:simple_toa})) & $t_1 \cdot (\epsilon_B - \epsilon_A)$ & RX & TX \\
			\hline
			\gls{twr} (\ref{sct:twr}) & $\frac{1}{2} (\epsilon_A - \epsilon_B) \cdot D_B$ & TX + RX & TX + RX\\
			\hline
			\gls{sds-twr} (\ref{sct:ds_twr}) & $\frac{1}{4} (\epsilon_A - \epsilon_B) \cdot (D_B - D_A)$ & TX + RX & TX + RX \\
			\hline
			Asym-DS-\gls{twr} (\ref{sct:asym_ds_twr}) & $\frac{1}{2}(\epsilon_A + \epsilon_B) \cdot d_{AB}$ & TX + RX & TX + RX \\
			\hline
			\hline
			Simple TDOA (\ref{sct:simple_tdoa}) & $(\epsilon_B - \epsilon_A) \cdot t_1 $ & RX & TX \\
			\hline
			Whistle (\ref{sct:whistle}) & $ (\epsilon_B - \epsilon_A) \cdot D_B$ & RX + one TX & TX \\
			\hline
			\gls{djkm} (\ref{sct:djkm}) & $ (\epsilon_T - \epsilon_B) \cdot D_B$ & RX + TX & RX \\
			\hline
			DP-Whistle (\ref{sct:dp_whistle}) & $(\epsilon_A + \epsilon_B) \cdot d_{AB}$ & RX + one TX & TX\\
			\hline
		\end{tabular}
		\label{tbl:error_comparision}
	\end{table*}

		Using the findings from \citet{dpw} the \gls{tdoa} value can then be calculated as follows:
		\begin{align}
			T_{AB}^S &:= \frac{R_A R_B - D_A D_B}{R_A + D_A} - d_{AB} \label{eqn:dpw_eqn_a}\\
					 &= \frac{R_A R_B - D_A D_B}{R_B + D_B} - d_{AB} \label{eqn:dpw_eqn_b}
		\end{align}
		Using $R_A + D_A = R_B + D_B$ we conclude:
		\begin{IEEEeqnarray}{rCl}
			T_{AB}^S = \frac{2(R_A R_B - D_A D_B)}{R_A + R_B + D_A + D_B} - d_{AB} \label{eqn:dpw_formula}
		\end{IEEEeqnarray}

		%\gls{dpw} has another interesting property which allows to increase error robustness and to develop an efficient protocol.
		%It is possible to reorder any of the signals in the scheme arbitrarily while still being able to apply the formula to it. 
		%For example \textbf{Fig. \ref{fig:dpw_shifted}} is still a valid form. This means that if only every second Tag signal is mirrored, the \gls{tdoa} values
    %can still be calculated with the second

		%\begin{figure}
		%	\centering
		%	\includegraphics{figures/schemes/dpw_shifted.pdf}
		%	\caption{Double Pulsed Whistle (shifted)}
		%	\label{fig:dpw_shifted}
		%\end{figure}

		% Now we compare the error between the other synchronization-free-\gls{tdoa} methods and \gls{dpw}.
		Again we use the same erroneous clock model as in \eqref{eqn:ds-twr_error_model} conforming to our model design in \eqref{eqn:error_modulation}
		to define the erronous timespans ($\hat{R}_A, \hat{D}_A, \hat{R}_B, \hat{D}_B$).
		% \begingroup
		% \allowdisplaybreaks
		% \begin{align*}
		% 	\hat{R}_A				&:= R_A \cdot (1 + \epsilon_A) \\
		% 	\hat{D}_A				&:= D_A \cdot (1 + \epsilon_A) \\
		% 	\hat{R}_B				&:= R_B \cdot (1 + \epsilon_B) \\
		% 	\hat{D}_B				&:= D_B \cdot (1 + \epsilon_B)
		% \end{align*}
		Then applying it to \eqref{eqn:dpw_eqn_a} we define:
		\begin{align*}
			\hat{T}_{AB}^S			&:= \frac{\hat{R}_A \hat{R}_B - \hat{D}_A \hat{D}_B}{\hat{R}_A + \hat{D}_A} - d_{AB}\\
		\end{align*}
		And equivalently using \eqref{eqn:dpw_eqn_b} we derive:
		\begin{align*}
										&= \frac{\hat{R}_A \hat{R}_B - \hat{D}_A \hat{D}_B}{\hat{R}_B + \hat{D}_B} - d_{AB}\\
		\end{align*}
		We transform the term with denominator $\hat{R}_B + \hat{D}_B$:
		\begin{align*}
			\hat{T}_{AB}^S			&= \frac{(1 + \epsilon_A)(1 + \epsilon_B)}{(1 + \epsilon_B)} \cdot \frac{R_A R_B - D_A D_B}{R_A + D_A} - d_{AB} \\
		\end{align*}
		Adding and subtracting $(1 + \epsilon_A)d_{AB}$ to the term brings:
		\begin{align*}
										=&\;(1 + \epsilon_A) \cdot \frac{R_A R_B - D_A D_B}{R_A + D_A}\\
										&- (1 + \epsilon_A)d_{AB} + (1 + \epsilon_A)d_{AB} - d_{AB} \\
		\end{align*}
		Then substituting with \eqref{eqn:dpw_eqn_a} delivers:
		\begin{align*}
										&= (1 + \epsilon_A) \cdot T_{AB}^S + (1 + \epsilon_A)d_{AB} - d_{AB} \\
										&= (1 + \epsilon_A) \cdot T_{AB}^S + \epsilon_Ad_{AB} \\
		\end{align*}
		The difference between the erroneous and error-free value is the error:
		\begin{align*}
			\tilde{T}_{AB}^S 		&:= \hat{T}_{AB}^S - T_{AB}^S \\
										&= \epsilon_AT_{AB}^S + \epsilon_Ad_{AB}\\
									  	&\leq 2\epsilon_Ad_{AB}\\
		\end{align*}
		The fact that $T_{AB}^S \leq d_{AB}$ justifies the last step here.
		That is true since the \gls{tdoa} ($T_{AB}^S$) can never grow larger than the distance between the measuring Anchor pair ($d_{AB}$).

		Equivalently we can apply this to the term with the divisor $\hat{R}_B + \hat{D}_B$:
		\begin{align*}
			\hat{T}_{AB}^S &= \frac{(1 + \epsilon_A)(1 + \epsilon_B)}{(1 + \epsilon_A)} \cdot \frac{R_A R_B - D_A D_B}{R_A + D_A} - d_{AB} \\
								&= (1 + \epsilon_B) \cdot T_{AB}^S + \epsilon_Bd_{AB} \\
			\tilde{T}_{AB}^S &= \epsilon_BT_{AB}^S + \epsilon_Bd_{AB} \\
								&\approx 2\epsilon_Bd_{AB}
		\end{align*}
		Combining the two fraction like in \eqref{eqn:dpw_formula} we end up with:
		\begin{IEEEeqnarray}{rCl}
			\tilde{T}_{AB}^S \approx (\epsilon_A + \epsilon_B)d_{AB} \label{eqn:dpw_error}
		\end{IEEEeqnarray}
		% \endgroup
		We observe that the error is now only depending on the distance value $d_{AB}$ which is in order of nanoseconds for typical positioning setups.
		We further multiply it with $\epsilon$'s which are in magnitudes of $\pm20$\gls{ppm}.
		That means that the error is in no significant magnitudes.

\section{Summary and Conclusion}\label{sct:comparision}
	In the Table \ref{tbl:error_comparision} all of the methods described in this paper are listed.
	Their worst-case clock-error and the required device capabilities for Anchors and Tags are listed for each method.
	For better comparison of the worst-case clock-drift errors the methods are divided into \gls{toa} and \gls{tdoa} methods.
	On the right, the interface requirements for Anchor- and Tag-Nodes are noted for each method.
	
  The table reiterates that all methods which use double-pulses like \gls{dpw} and \gls{asym-ds-twr} are much more robust against clock errors. Especially for 
  radio based \gls{tof} methods like UWB it is important to reduce these error sources because the contributions of the clock error in Whistle and \gls{djkm} 
  could significantly distort the result. 

  %Similarities and relationships between the methods (for instance the Double-Pulse) were highlighted (e.g., \gls{dpw} and \gls{djkm}, \gls{dpw} and \gls{asym-ds-twr}).
  %The observations we made about those similarities and the strengths and weaknesses of certain methods provide a deeper understanding about the toolset 
  %which is available to us
  %when designing or implementing an \gls{ftm}-positioning protocol.
    
  In future work, approaches like \gls{djkm} can be extended to incorporate the Double-Pulse similar to 
  the transition from Whistle to \gls{dpw}. This would keep the advantages of \gls{djkm} like allowing unlimited number of Tags with the advantages of high 
  robustness against clock errors.

  This approach is related to our efforts in designing Double Pulsed Positioning \cite{dpp_paper} a novel infrastructure-less and synchronisation-free \gls{tof} method.
	%In this paper, the fundamental techniques and principles of \gls{toa} and \gls{tdoa} measurement were explained and analyzed.
	%Each methods advantages and disadvantages, and its sources of errors were analyzed.
	%The analysis was conducted using a clock-error model and analysis techniques inspired by typical works in the field of \gls{ftm}-positioning (e.g., \cite{dw_manual}, \cite{alternative_twr}, \cite{dpw}).
%

	% In the following Chapter that understanding (and toolset) is used to construct a novel improved \gls{ftm}-positioning-protocol.

% \section{Future Work}
% Lorem ipsum dolor sit amet, consectetuer adipiscing
% elit. Etiam lobortis facilisis sem. Nullam nec mi et neque
% pharetra sollicitudin. Praesent imperdiet mi nec ante. Donec
% ullamcorper, felis non sodales commodo, lectus velit ultrices
% augue, a dignissim nibh lectus placerat pede.

% \section*{Acknowledgment}
% 	The authors would like to thank all reviewers for their support improving this document.

\bibliography{analysis}

\end{document}